\documentclass[10pt,twocolumn,letterpaper]{article}
\usepackage{cvpr}
\usepackage{times}
\usepackage{epsfig}
\usepackage{graphicx}
\usepackage{amsmath}
\usepackage{amssymb}
\usepackage{booktabs}
\usepackage{multirow}
\usepackage{xcolor}
\usepackage[breaklinks,colorlinks]{hyperref}
\usepackage{amsthm}
\usepackage{stfloats}
\usepackage{enumitem}

\usepackage{tikz}

\newcommand{\ASVspoof}[1]{\textit{ASVspoof #1}}

\begin{document}



\title{Probing Speaker Identity Sensitivity in Audio Deepfake Detectors}
\author{Daniyal Kabir Dar and Arun Ross\\
Michigan State University, USA\\
{\tt\small \{dardaniy, rossarun\}@msu.edu}
}

\maketitle

\begin{tikzpicture}[remember picture,overlay]
\node[
    anchor=north,
    align=center,
    text width=0.92\paperwidth
] at ([yshift=-0.35in]current page.north) {%
    \small\textcolor{red}{\textbf{%
    This paper has been accepted for publication at the
    IEEE/IAPR International Joint Conference on Biometrics (IJCB),\\
    Rome, Italy, September 2026.}}
};
\end{tikzpicture}

\thispagestyle{empty}
\thispagestyle{empty}

\begin{figure*}[bt]
  \centering
  \includegraphics[width=0.92\textwidth,
  trim=10 10 10 10, clip]{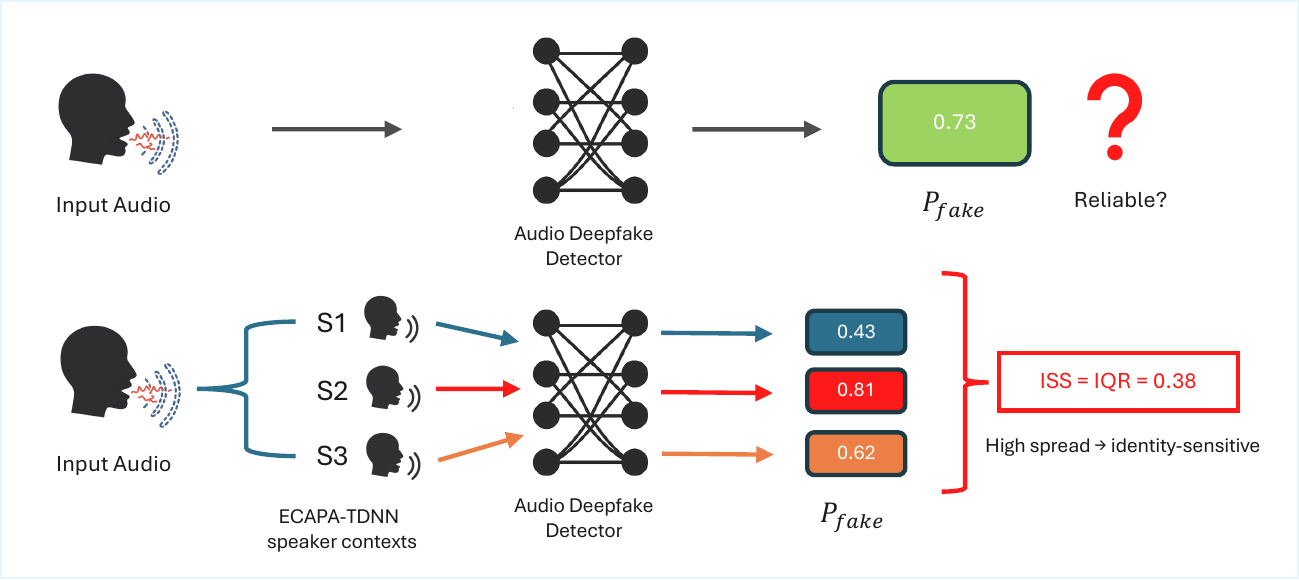}
  
\caption{\textbf{The ISS concept.}
\textit{Top:} A standard detector produces a
single score $p_{\text{fake}}$ with no indication
of whether the prediction is reliable ---
the same score could reflect synthesis artifacts
or identity-dependent behavior.
\textit{Bottom:} ISS evaluates the same audio
under $K$ varied speaker identity contexts
($S_1$, $S_2$, $S_3$) via ECAPA-TDNN prototypes
and measures the score spread via IQR (interquartile range)\@.
A high spread (ISS $= 0.38$) indicates an
identity-sensitive decision; low spread indicates
stability across speaker contexts.
No ground-truth labels are required at inference time.}
  \label{fig:concept}
\end{figure*}

\begin{abstract}
Audio deepfake detectors are trained to distinguish genuine speech 
from synthetic speech and often perform well on  standard benchmarks. Yet the same detector that achieves less than 1\% error on one 
dataset can see its error rate increase twentyfold when evaluated on a different dataset. We argue that one contributing factor is 
speaker-identity reliance: standard training corpora correlate speaker 
identity with the genuine/synthetic label, allowing detectors to partially 
rely on speaker-related cues rather than synthesis artifacts alone.
We propose the Identity Sensitivity Score (ISS), a per-utterance 
diagnostic that quantifies how much a detector's output changes across different speaker identity contexts. ISS requires 
no ground-truth labels at inference time and can be computed from the 
detector score and a pool of reference speaker examples. Across two detectors 
and two datasets, incorrectly classified utterances have ISS scores 29 
to 52 times higher than correctly classified utterances, and ISS alone 
predicts misclassification with area-under-curve (AUC) up to 0.954. To test whether ISS actually
captures identity-sensitive behavior rather than serving only as a proxy 
for prediction confidence, we apply voice conversion to 500 utterances 
and measure the resulting detector-score shift. Utterances flagged as 
identity-sensitive by ISS respond 19 to 30 times more strongly to this 
manipulation than utterances flagged as stable. These results position 
ISS as a practical inference-time diagnostic for speaker-dependent 
failure analysis in audio deepfake detection.
\end{abstract}

\section{Introduction}
\label{sec:intro}

Modern deepfake systems either synthesize speech
directly from text using text-to-speech
(TTS)~\cite{tan2021survey} or convert an existing
recording to sound like someone else using voice
conversion (VC)~\cite{sisman2020overview}, while
preserving the original words spoken.
These two paradigms have matured
rapidly~\cite{sisman2020overview}, raising serious
challenges for speaker verification~\cite{khanjani2023audio}, forensic
authentication~\cite{cheng2025forensic}, and voice-based communication~\cite{khanjani2023audio}.

This has led to the development of audio deepfake
{\em detectors} that are essentially classifiers
that input an audio sample and output a score in
the range $[0,1]$; a $0$ indicates a real sample
and a $1$ indicates a synthetic sample.
These detectors achieve low error rates on standard benchmarks: AASIST~\cite{jung2022aasist}
achieves sub-1\% Equal Error Rate (EER) on
\ASVspoof{2019 LA}~~\cite{wang2020asvspoof}, while
RawNet2~\cite{tak2021end} achieves 4.11\%.
Yet AASIST's EER increases 21$\times$
(0.83\%~$\to$~17.39\%) from \ASVspoof{2019 LA} to \ASVspoof{2021 LA} ~\cite{delgado2021asvspoof,muller2022does}.
While this gap can be attributed to recording
conditions, telephone codec artifacts, and speaker
mismatch, these do not fully account for why
detectors fail in the specific ways they do.

We investigate one underexplored contributor:
detectors may rely on speaker identity as a
predictive cue rather than solely on synthesis artifacts
(Figure~\ref{fig:concept}).
In standard benchmarks, genuine recordings always
come from real speakers while synthetic ones
always come from synthesis systems (e.g.,
\ASVspoof{2019 LA} contains genuine speech from
real speakers and spoofed speech from
TTS/VC attack systems~\cite{wang2020asvspoof,
tan2021survey}), creating a
systematic correlation between speaker identity
and genuine/synthetic label throughout training.
A detector that exploits this correlation of
associating certain speaker characteristics with
the genuine label and others with the synthetic
label performs well on speakers seen during
training but fails when new speakers and acoustic
conditions are encountered, aggregate
metrics like EER cannot reveal  these correlations.

We introduce the Identity Sensitivity Score (ISS) that measures how a deepfake detector's output score changes when we vary the identity of the input audio sample.  
We evaluate ISS on \ASVspoof{2019} and
\ASVspoof{2021} LA (Logical Access)
using two architecturally distinct detectors:
AASIST (graph attention network) and RawNet2
(end-to-end CNN on raw waveforms).
Our findings build a consistent picture.
\emph{First}, incorrectly classified utterances
have 29 times higher ISS than correct ones for
AASIST and 52 times for RawNet2, with ISS alone
predicting misclassification at an AUC of 0.954
(Figure~\ref{fig:roc}).
\emph{Second}, controlled voice conversion experiments support the identity-sensitivity interpretation, with high-ISS utterances responding
19.2 times more strongly for AASIST and
30.7 times for RawNet2 (both $p < 10^{-70}$).
\emph{Third}, when AASIST is evaluated on
\ASVspoof{2021 LA} instead of \ASVspoof{2019 LA},
its EER increases 21$\times$
(0.83\%~$\to$~17.39\%).
Over the same shift, the ISS separation ratio
between incorrect and correct predictions
amplifies 24$\times$ (29$\times \to$ 690$\times$),
suggesting that the predictions which fail under
distribution shift are disproportionately the
same ones that were identity-sensitive in-domain.
\emph{Fourth}, incorrect
predictions on RawNet2 hybrid attacks (combined
TTS and VC) show no ISS elevation, indicating
acoustic rather than identity-driven failures,
while TTS and VC attack errors show strong ISS
separation across both detectors.

\noindent\textbf{Contributions.}

\begin{itemize}[leftmargin=*, itemsep=0pt,
topsep=1pt]
  \item We introduce ISS, a
  per-utterance, label-free metric for
  quantifying speaker identity sensitivity
  in audio deepfake detection, grounded in
  the sensitivity of sigmoid scores near
  the decision boundary.
  \item We conduct controlled voice conversion
  experiments and evaluate ISS
  across two detectors, two datasets, and
  three attack categories with zero-shot
  evaluation on \ASVspoof{2021 LA}.
\item We perform ablation studies confirming ISS stability
across enrollment configurations and determining the
perturbation strength using a labeled development set.
\end{itemize}


\section{Related Work}
\label{sec:related}

\subsection{Audio Deepfake Detection}
\label{sec:related_add}

Audio deepfake detectors are trained to classify an input audio sample or utterance as either
genuine or synthetic, and are primarily evaluated on ASVspoof-style
benchmarks under the Logical Access scenario, where attacks are generated
by text-to-speech and voice conversion systems~\cite{todisco2019asvspoof,
wang2020asvspoof, liu2023asvspoof2021}. Broader surveys and challenge
reports show that, despite rapid progress, generalization to unseen attacks,
recording conditions, and speaker populations remains a central open
problem~\cite{yi2023audio, yi2022add, yi2023add2023}.

Architecturally, these systems range from models that operate on handcrafted
spectral features to end-to-end networks that learn directly from raw
waveforms~\cite{tak2021end, jung2022aasist}. More recent work has
incorporated graph attention, local-global dependency modeling, state-space
sequence modeling, and self-supervised speech representations to improve
robustness across spoofing conditions~\cite{tak2021spectro,
liu2023leveraging, chen2024rawbmamba, zhang2024audio, tak2022automatic}.
Additional datasets such as \textit{WaveFake} have further expanded evaluation beyond
the original ASVspoof setting~\cite{frank2021wavefake}.

Despite this progress, cross-dataset generalization remains poor
~\cite{muller2022does, liu2023asvspoof2021}. The causes are typically
attributed to channel variability, codec artifacts, attack mismatch, and
speaker mismatch between training and test conditions. One factor that is relatively less explored is
the impact of speaker identity.
For example, in \ASVspoof{2019 LA}, the synthetic
audio samples were generated by TTS and VC systems
trained on the real speakers present in the
dataset~\cite{wang2020asvspoof}
(see Section~\ref{sec:datasets}).
This creates a systematic correlation between
speaker identity and the genuine/synthetic labels
even in the training data. Phonetic content has been
shown to modulate how well speaker identity is preserved
in adversarial voice attacks~\cite{dar2026impact},
and prior work has demonstrated that speaker
identity is exploitable in deepfake detection, both as an explicit
feature~\cite{pan2022speaker, gu2024utilizing} and as the sole basis for a
generalizable detector~\cite{pianese2022deepfake}. Whether standard detectors
learn to exploit this correlation implicitly, and whether doing so contributes
to generalization failure, remains an open question.

\subsection{Bias and Fairness}
\label{sec:related_bias}

In audio deepfake detection, identity-induced disparities 
have received little systematic attention. Kawa 
et al.~\cite{kawa2023improved} observed that detector 
performance varies across speaker groups, but their 
analysis operated at the group level and did not provide 
a way to diagnose individual predictions. What is missing 
is a per-utterance tool that can attribute a specific 
detector's prediction behavior to speaker identity, enabling the 
kind of targeted diagnosis that aggregate statistics 
cannot provide.

\subsection{Uncertainty Quantification}
\label{sec:related_uq}

Standard uncertainty quantification methods such as 
Monte Carlo dropout~\cite{gal2016dropout} and deep 
ensembles~\cite{lakshminarayanan2017simple} estimate 
how confident a model is in its predictions, but do not 
explain where that uncertainty comes from. Score-based 
alternatives, including entropy, margin, and prediction 
confidence, are simpler but face a fundamental 
limitation: for a binary classifier with a single 
sigmoid output, all three reduce to the same monotonic 
function of $|p_{\text{fake}} - 0.5|$, where
$p_{\text{fake}} \in [0,1]$ is the detector's
estimated probability that the input is synthetic.
Each depends only on the distance from
the decision boundary at $p_{\text{fake}} = 0.5$: all
three are maximized there and decrease symmetrically
as $p_{\text{fake}}$ moves toward 0 or 1. Consequently,
they produce identical AUC values when used as error
predictors (Table~\ref{tab:baselines}).
They measure how 
close a prediction is to the decision boundary, but they 
cannot say why it is close. Two predictions with 
identical scores could reflect genuinely ambiguous 
synthesis artifacts, or a decision that flips depending 
on which speaker is present. From a score-based 
perspective, these are indistinguishable.

ISS addresses this directly. Rather than asking how 
uncertain a prediction is, it asks how much of that 
uncertainty is attributable to speaker identity 
specifically. This makes it complementary to existing 
uncertainty measures rather than a replacement for them, 
and it opens up a kind of per-utterance diagnostic that 
score-based metrics cannot provide.

\subsection{Voice Conversion for Controlled Evaluation}
\label{sec:related_vc}

Voice conversion systems transfer the vocal 
characteristics of a target speaker onto a source 
recording while preserving its linguistic 
content~\cite{sisman2020overview}. In the deepfake 
detection literature, voice conversion has been used 
primarily to augment training data~\cite{yi2022add} and 
to test how detector performance varies across different 
speaker characteristics~\cite{kawa2023improved}. In both 
cases it is used as a data generation tool rather than 
an analytical one.

We use voice conversion differently, as a controlled 
intervention for validating ISS. If ISS correctly 
identifies utterances whose detector scores are sensitive 
to speaker identity, then changing the speaker 
identity of those utterances via voice conversion should 
produce larger shifts in detector output than doing the 
same for utterances which ISS flags as stable. This gives us a 
way to verify that ISS is measuring genuine 
identity-sensitive behavior rather than serving as a 
proxy for prediction confidence. We use 
FreeVC~\cite{li2023freevc} for this purpose, which 
enables high-quality one-shot voice conversion without 
requiring a transcript of the source audio.
\section{Method}
\label{sec:method}

\subsection{Problem Formulation}
\label{sec:formulation}

Let $A$ be an audio utterance and $D(\cdot)$ a
deepfake detector that outputs
$p_{\text{fake}}(A) \in [0,1]$.
Standard evaluation asks: what is the aggregate
error rate over a labeled test set?
We ask a different question: for a specific
utterance $A$, how much does
$p_{\text{fake}}$ change when we vary the speaker
identity?
A detector that relies heavily on speaker identity
will show large score changes under identity
variation; a detector relying on synthesis-specific
artifacts will show small changes.
This question can be answered without ground-truth
labels, enabling deployment-time diagnostics.



\subsection{Speaker Identity Encoding}
\label{sec:encoding}

We use ECAPA-TDNN~\cite{desplanques2020ecapa}
pretrained on VoxCeleb~\cite{nagrani2017voxceleb}, to extract $d$-dimensional
speaker embeddings.
For utterance $A$, let
$\mathbf{e}(A) \in \mathbb{R}^d$ be its
L2-normalized embedding.
For enrolled speaker $S_i$, the prototype is the
mean embedding over $N$ enrollment utterances:
\begin{equation}
  \mathbf{E}(S_i) = \frac{1}{N}
  \sum_{j=1}^{N} \mathbf{e}(u_j^{(i)})
  \label{eq:prototype}
\end{equation}
Cosine similarity measures identity proximity:
\begin{equation}
  \text{cos}(A, S_i) =
  \mathbf{e}(A)^\top \mathbf{E}(S_i)
  \label{eq:cosine}
\end{equation}
Enrollment prototypes
$\mathbf{E}(S_i)$ are computed once offline and
cached; at inference, ISS requires only a single
ECAPA-TDNN forward pass on the query utterance.

\subsection{Identity-Conditioned Score}
\label{sec:conditioned}

We inject a speaker identity context into the
detector score via logit-space perturbation.
Let $\ell(A) = \log\!\left(p_{\text{fake}}(A) /
(1 - p_{\text{fake}}(A))\right)$ be the log-odds
of the base detector score.
The identity-conditioned score is:
\begin{equation}
  p_{\text{fake}}(A \mid S_i) =
  \sigma\!\left(\ell(A) +
  \alpha \cdot \text{cos}(A, S_i)\right)
  \label{eq:conditioned}
\end{equation}
where, $\sigma(x) = 1/(1+e^{-x})$ and $\alpha \geq 0$
controls perturbation strength.
Critically, the audio itself is never modified;
only the identity context used in detection score computation
varies.
This is a logit-space sensitivity probe rather
than a direct detector-internal intervention.
Addition in logit
space corresponds to multiplication in odds space,
consistent with Bayesian score
fusion~\cite{brummer2006application}.

\subsection{Identity Sensitivity Score}
\label{sec:iss}

Given query audio $A$, we sample $K$ alternative
speakers $\{S_1, \ldots, S_K\}$ from an enrollment
pool (excluding $A$'s true speaker), compute $K$
identity-conditioned scores, and define ISS as
their spread via interquartile range (IQR):
\begin{equation}
  \text{ISS}(A) = \text{IQR}
  \left\{p_{\text{fake}}(A \mid S_i)\right\}_{i=1}^{K}
  \label{eq:iss}
\end{equation}
IQR ($Q_3 - Q_1$, the difference between the 75th and
25th percentiles) is used rather than variance
because it is robust to acoustically extreme
outlier speakers.
At inference, ISS reuses
$p_{\text{fake}}(A)$ plus one ECAPA-TDNN forward
pass; no audio is generated and no extra detector
passes are required.
ISS is label-free \emph{at inference time}: it
requires only the audio and enrollment prototypes,
with no ground-truth labels needed for deployment.
However, note that tuning $\alpha$ and computing enrollment
prototypes does require a small labeled development
set, as described in Section~\ref{sec:iss_config}.



\subsection{Theoretical Motivation}
\label{sec:theory}

ISS is geometrically coupled to decision-boundary
proximity through the slope of the sigmoid.
For a fixed identity perturbation
$\delta_i = \alpha \cdot \cos(A, S_i)$,
the probability change
$\sigma(\ell(A)+\delta_i)-\sigma(\ell(A))$
is largest when the perturbed logits lie near
the decision boundary, where the sigmoid
derivative $\sigma'(x)=\sigma(x)(1-\sigma(x))$
is maximized.
This explains why high-ISS predictions often
coincide with errors: both tend to occur near
sensitive regions of the decision function.
However, ISS provides information that entropy,
margin, and confidence cannot: the
identity-specific component of this boundary
proximity.
These standard uncertainty measures depend only
on the base score $p_{\text{fake}}(A)$, whereas
ISS depends on how that score varies across
speaker contexts.
Two utterances with identical
$p_{\text{fake}}(A) = 0.51$ will have
identical entropy, margin, and confidence, but
may have very different ISS: one may be stably
uncertain across speakers, while the other may
be uncertain because its score is
identity-dependent.

\subsection{ISS as a Deployment Diagnostic}
\label{sec:diagnostic}

ISS enables three label-free
applications: \textbf{confidence flagging}
(utterances above threshold $\tau$ flagged for
review; AUC up to 0.954), \textbf{sensitivity
auditing} (stratified speaker analysis identifies
high-risk subpopulations), and \textbf{shift
monitoring} (ISS amplification tracks distribution
change; 21 times EER increase corresponds to
24 times ISS amplification for AASIST).

\section{Experimental Setup}
\label{sec:setup}

\subsection{Datasets}
\label{sec:datasets}

\noindent\textbf{ASVspoof 2019 LA.}
The evaluation set contains 71,237 utterances:
7,355 bonafide (10.33\%) and 63,882 spoofed
(89.67\%) from 119 speakers, produced by 13 attack
systems (A07--A19) covering TTS, VC, and hybrid
synthesis methods~\cite{wang2020asvspoof}.

\noindent\textbf{ASVspoof 2021 LA.}
The evaluation set contains 181,566 utterances
(18,452 bonafide, 163,114 spoofed) from 48 speakers,
recorded under telephone channel and codec conditions
absent from the 2019 training
distribution~\cite{delgado2021asvspoof}.
No retraining or fine-tuning is performed on 2021
data; all models are evaluated zero-shot, making
this a strict test of generalization.

\subsection{Deepfake Detectors}
\label{sec:detectors}

\noindent\textbf{AASIST.}
We use the official pretrained checkpoint
(297,866 parameters)~\cite{jung2022aasist},
achieving 0.83\% EER on \textit{2019 LA}.
Detector output is a 2-dimensional logit vector;
we apply softmax and take the spoof class
probability as $p_{\text{fake}}(A) \in [0,1]$.

\noindent\textbf{RawNet2.}
RawNet2~\cite{tak2021end} extends the original
RawNet architecture~\cite{jung2019rawnet} with
sinc-based filters operating directly on raw
waveforms. We train from scratch using the official ASVspoof
2021 baseline~\cite{asvspoof2021baseline} for 100
epochs on the 2019 LA training partition.
Our checkpoint selected by
validation EER, achieves 4.11\% EER on \textit{2019 LA}
and 12.41\% on \textit{2021 LA}.
These two detectors --- one graph attention network,
one end-to-end CNN --- allow us to assess whether
identity-sensitive behavior is architecture-specific
or systemic.

\subsection{ISS Configuration}
\label{sec:iss_config}

\noindent\textbf{Speaker embeddings.}
We use ECAPA-TDNN~\cite{desplanques2020ecapa}
via SpeechBrain~\cite{ravanelli2021speechbrain}, extracting
192-dimensional L2-normalized embeddings.
This is an external speaker model; ISS does not
access the internal representations of the
deepfake detector being probed.

\noindent\textbf{Enrollment prototypes.}
Speaker prototypes are built from all 20 speakers
in the \ASVspoof{} \textit{2019 LA} training set --- the
complete set of available training speakers, as
the \ASVspoof{} \textit{2019 LA} training partition
contains exactly 20 speakers (8 male, 12
female)~\cite{wang2020asvspoof} --- using the
mean of $N{=}5$ enrollment utterances per speaker.

\noindent\textbf{Hyperparameters.}
We fix $K{=}10$ alternative speakers per utterance
and perturbation strength $\alpha{=}5.0$.
$K$ is selected based on ablation showing stable
ISS ratios of 29 to 32 for
$K \in \{3, 5, 10, 20\}$
(Section~\ref{sec:ablation}).
$\alpha$ is tuned on the \textit{2019 LA} development set
by maximizing the correct/incorrect ISS ratio,
selecting the largest value before bonafide/spoof
ordering inverts (Section~\ref{sec:ablation}).
Concretely, $\alpha{=}5$ preserves
the bonafide/spoof median ordering on the
development set, while $\alpha{=}10$ inverts it
(Table~\ref{tab:ablation_alpha} footnote), so we
select $\alpha{=}5$ as the largest valid value.
We note that $\alpha$ tuning requires a small
labeled development set; this is a one-time
offline calibration step, not a deployment-time
label requirement.

\subsection{Voice Conversion Validation Protocol}
\label{sec:vc_protocol}

To probe whether ISS-predicted identity sensitivity
corresponds to actual detector sensitivity to
identity change, we perform a controlled
manipulation experiment.

From \textit{ASVspoof 2019 LA}, we select 500 bonafide
utterances per detector, all correctly classified
($p_{\text{fake}} < 0.5$, $y_{\text{true}}{=}0$):
the 250 utterances with the highest ISS and the
250 with the lowest ISS from the correctly
classified pool.
For AASIST, this rank-based selection corresponds
to ISS $> 0.05$ (high group) and ISS $< 0.01$
(low group).
For RawNet2, whose ISS distribution is
substantially more compressed, the high group
has median $3.1 \times 10^{-3}$ and the low
group has median $5.8 \times 10^{-6}$;
absolute thresholds are not used to ensure
comparable group sizes across detectors.

We apply FreeVC~\cite{li2023freevc} to transfer
a randomly selected target speaker's
characteristics while preserving linguistic
content, producing
$A_{\text{conv}} =
\text{FreeVC}(A_{\text{src}}, A_{\text{tgt}})$.
We measure
$\Delta\text{score} =
|p_{\text{fake}}(A_{\text{conv}}) -
p_{\text{fake}}(A_{\text{src}})|$
and test whether the high-ISS group shows a
larger response than the low-ISS group via
Mann-Whitney U test and Cohen's $d$.

\subsection{Evaluation Metrics}
\label{sec:metrics}

\noindent\textbf{Detection performance.}
EER ($\downarrow$) and AUC ($\uparrow$) following
standard ASVspoof protocol~\cite{wang2020asvspoof}.

\noindent\textbf{ISS diagnostics.}
We report: (i)~median ISS for correctly and
incorrectly classified utterances and their ratio;
(ii)~error-curve AUC (ISS as a binary predictor
of misclassification); (iii)~Pearson correlation,
$r$, between ISS and error label.
All ISS results use $K{=}10$, $\alpha{=}5.0$.

\noindent\textbf{Baseline comparison.}
We compare ISS against entropy
$H(p) {=} {-}p\log p {-} (1{-}p)\log(1{-}p)$,
margin $= 1 - 2|p_{\text{fake}} - 0.5|$,
and confidence
$= 1 - \max(p_{\text{fake}}, 1-p_{\text{fake}})$.
For binary classifiers with a single sigmoid
output, these are mathematically equivalent ---
all monotonic transformations of
$|p_{\text{fake}} - 0.5|$ --- and represent a
single baseline family.

\noindent\textbf{Statistical significance.}
All group comparisons use the one-sided
Mann-Whitney U test. Wherever p-values underflow
double-precision floating point, we report
$p \approx 0$.

\section{Results}
\label{sec:results}

\subsection{Baseline Detection Performance}
\label{sec:baseline_perf}

Table~\ref{tab:detector_perf} confirms the
generalization gap motivating this work.
AASIST achieves 0.83\% EER on \textit{2019 LA} but degrades
to 17.39\% on \textit{2021 LA} --- a 21 times increase
attributable to telephone channel and codec
conditions absent from training.
RawNet2 degrades more modestly (4.11\%~$\to$~12.41\%,
3 times) from an already higher baseline.
The divergent degradation patterns allow us to
assess whether ISS behavior under shift is
architecture-specific or universal.

\begin{table}[t]
\centering
\caption{Detector performance. The 21 times EER
increase for AASIST illustrates the generalization
gap that motivates ISS.}
\label{tab:detector_perf}
\setlength{\tabcolsep}{3pt}
\begin{tabular}{llcccc}
\toprule
Dataset & Detector & EER (\%)$\downarrow$ & AUC$\uparrow$
& Correct & Incorrect \\
\midrule
2019 LA & AASIST  & 0.83  & 0.9993 & 68,949  & 2,288  \\
2019 LA & RawNet2 & 4.11  & 0.9923 & 66,014  & 5,223  \\
2021 LA & AASIST  & 17.39 & 0.9175 & 146,307 & 35,259 \\
2021 LA & RawNet2 & 12.41 & 0.9317 & 169,231 & 12,335 \\
\bottomrule
\end{tabular}
\end{table}

\subsection{ISS Separates Correct from Incorrect
Predictions}
\label{sec:iss_separation}

\begin{figure}[t]
  \centering
  \includegraphics[width=\columnwidth]{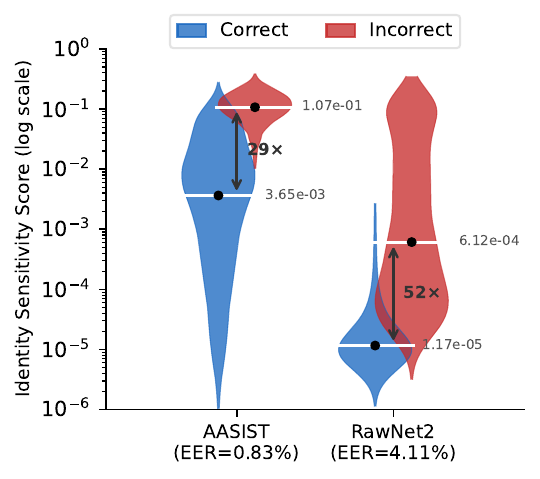}
  \caption{
    ISS distributions for correctly and incorrectly
    classified utterances on \textit{ASVspoof 2019 LA}
    (log scale, $K{=}10$, $\alpha{=}5.0$).
    The separation --- 29 times for AASIST and
    52 times for RawNet2 ($p \approx 0$) ---
    indicates that identity-sensitive predictions
    are substantially more likely to be errors.
  }
  \label{fig:violin}
\end{figure}

Table~\ref{tab:iss_main} and Figure~\ref{fig:violin}
show that ISS is substantially higher for
incorrectly classified utterances than correct ones.
On \textit{ASVspoof 2019 LA}, the median ISS is 29 times
higher for incorrect predictions than correct ones
for AASIST (0.107 vs.\ 0.004), and 52 times
higher for RawNet2
(6.12$\times 10^{-4}$ vs.\ 1.17$\times 10^{-5}$).
This pattern is consistent across two architecturally
distinct detectors, suggesting that
identity-sensitive behavior is a general property
rather than an artifact of any specific model.

Under cross-dataset evaluation on \textit{2021 LA}, the
picture becomes more informative.
AASIST's ISS ratio amplifies from 29 times to
690 times --- a 24 times amplification as EER
increases 21 times --- consistent with the
hypothesis that identity-sensitive behavior
concentrates failure under acoustic distribution
shift.
RawNet2's ratio moderates from 52 times to 24 times;
under shift, its errors are more broadly distributed
rather than concentrated at identity-sensitive
boundaries.
In both cases ISS maintains strong separation,
remaining a meaningful predictor of errors under
distribution shift.

\begin{table}[t]
\centering
\caption{ISS statistics ($K{=}10$, $\alpha{=}5.0$).
ISS$_{\text{cor}}$ and ISS$_{\text{inc}}$ denote
median ISS for correct and incorrect predictions.
Error-curve AUC measures ISS as a label-free
predictor of misclassification.}
\label{tab:iss_main}
\setlength{\tabcolsep}{3pt}
\begin{tabular}{llccccc}
\toprule
Dataset & Det. &
ISS$_{\text{cor}}$ &
ISS$_{\text{inc}}$ &
Ratio &
AUC &
$r$ \\
\midrule
2019 LA & AASIST  & 3.65e-3 & 1.07e-1 & 29$\times$  & 0.954 & 0.506 \\
2019 LA & RawNet2 & 1.17e-5 & 6.12e-4 & 52$\times$  & 0.918 & 0.335 \\
2021 LA & AASIST  & 1.22e-5 & 8.44e-3 & 690$\times$ & 0.843 & 0.359 \\
2021 LA & RawNet2 & 9.51e-6 & 2.27e-4 & 24$\times$  & 0.855 & 0.275 \\
\bottomrule
\end{tabular}
\end{table}

\subsection{ISS Predicts Misclassification Without
Labels}
\label{sec:iss_prediction}

\begin{figure}[t]
  \centering
  \includegraphics[width=\columnwidth]{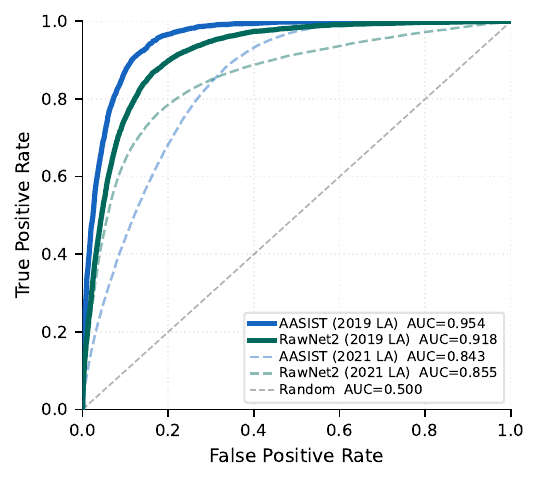}
  \caption{
    ROC curves for ISS as a label-free predictor
    of misclassification. ISS achieves AUC 0.954
    on AASIST (2019 LA) and maintains AUC above
    0.84 under distribution shift (2021 LA).
  }
  \label{fig:roc}
\end{figure}

ISS is not a deepfake classifier:
it is a label-free error predictor computed after
the detector decides.
Figure~\ref{fig:roc} shows that ISS is a reliable
predictor of misclassification without requiring
any labels.
ISS achieves AUC 0.954 for AASIST and 0.918 for
RawNet2 on 2019 LA: in 95.4\% and 91.8\% of
randomly drawn (correct, incorrect) pairs,
respectively, the incorrect prediction has higher
ISS.
Under distribution shift on \textit{2021 LA}, AUC decreases
to 0.843 (AASIST) and 0.855 (RawNet2), both
well above the score-based baselines
in all conditions.

\subsection{ISS vs.\ Uncertainty Baselines}
\label{sec:baselines}

Table~\ref{tab:baselines} compares ISS against
entropy, margin, and confidence, with three findings. First, ISS outperforms all baselines as an error
predictor across every detector-dataset combination
($p \approx 0$).
Second, all three baselines produce identical
AUC values --- confirming they are a single
baseline family, not three independent methods.
Third, the ISS advantage grows under shift:
from 1.8~pp on AASIST 2019 to 3.0~pp on
AASIST 2021, while baselines degrade alongside
the detector.

The deeper distinction is qualitative, not just
quantitative.
Entropy, margin, and confidence all answer the
same question: how close is this prediction to
the decision boundary?
ISS answers a different question: how much of
that boundary proximity is driven by speaker
identity?
Consider two utterances both with
$p_{\text{fake}} = 0.51$ --- identical entropy,
margin, and confidence.
If one has high ISS and the other low ISS, they
represent fundamentally different failure risks:
the high-ISS utterance's score is sensitive to
which speaker is present, while the low-ISS
utterance is stably uncertain regardless of
speaker context.
The attack-type results (Section~\ref{sec:attack_type}) 
demonstrate this distinction empirically: RawNet2 hybrid 
errors show no ISS elevation over correct predictions 
(ratio 0.2$\times$, $p = 1.00$), correctly identified by 
ISS as non-identity-driven, while entropy flags them 
identically to all other errors.

\begin{table}[t]
\centering
\caption{Error-curve AUC for ISS vs.\ score-based
uncertainty baselines. Entropy, margin, and
confidence are mathematically equivalent for
binary classifiers and shown as a single row.
All differences $p \approx 0$.}
\label{tab:baselines}
\setlength{\tabcolsep}{4pt}
\begin{tabular}{lcccc}
\toprule
Method &
\shortstack{RN2\\2019} &
\shortstack{RN2\\2021} &
\shortstack{AASIST\\2019} &
\shortstack{AASIST\\2021} \\
\midrule
\textbf{ISS (ours)} &
  \textbf{0.918} & \textbf{0.855} &
  \textbf{0.954} & \textbf{0.843} \\
Entropy/Margin/Conf. &
  0.865 & 0.825 & 0.936 & 0.813 \\
\midrule
\textit{Advantage} &
  \textit{+5.3} & \textit{+3.0} &
  \textit{+1.8} & \textit{+3.0} \\
\bottomrule
\end{tabular}
\end{table}

\subsection{Voice Conversion Validation}
\label{sec:vc}

The ISS--error correlation could be spurious:
ISS might simply track boundary proximity, with
the identity component coincidental.
To probe this, we intervene on speaker
identity via voice conversion on 500 correctly classified bonafide
utterances per detector (250 high-ISS, 250 low-ISS;
see Section~\ref{sec:vc_protocol}) and test whether
ISS predicts the magnitude of the detector's
response.

\noindent\textbf{AASIST.}
High-ISS utterances show mean
$\Delta\text{score} = 0.1069$ after FreeVC
conversion, compared to $0.0056$ for low-ISS
utterances --- a 19.2 times differential
(Mann-Whitney U $p = 6.76 \times 10^{-81}$,
Cohen's $d = 1.456$, Spearman $\rho = 0.846$).

\noindent\textbf{RawNet2.}
High-ISS utterances show mean
$\Delta\text{score} = 0.0401$, compared to $0.0013$
for low-ISS utterances --- a 30.7 times
differential (Mann-Whitney U
$p = 5.66 \times 10^{-72}$, Cohen's $d = 0.471$,
Spearman $\rho = 0.806$).

Both detectors show highly significant differential
response ($p < 10^{-70}$).
The consistent pattern --- high-ISS utterances
respond substantially more to identity manipulation,
while low-ISS utterances show minimal response ---
provides evidence that ISS is indeed measuring an
identity-sensitive component of detector behavior,
not merely boundary proximity.
AASIST's larger Cohen's $d$ (1.456 vs.\ 0.471)
reflects a more concentrated effect within its
score distribution; RawNet2's higher mean ratio
(30.7 times vs.\ 19.2 times) is consistent
with its higher in-domain ISS separation
(52 times vs.\ 29 times).

\begin{table}[t]
\centering
\caption{Voice conversion validation using FreeVC.
Per detector: top 250 ISS utterances (high-ISS)
and bottom 250 ISS utterances (low-ISS) from
correctly classified bonafide pool.
$\Delta$score $= |p_{\text{fake}}(A_{\text{conv}})
- p_{\text{fake}}(A_{\text{src}})|$.
Both differentials significant at $p < 10^{-70}$.}
\label{tab:vc}
\setlength{\tabcolsep}{4pt}
\begin{tabular}{lcccc}
\toprule
& \multicolumn{2}{c}{AASIST}
& \multicolumn{2}{c}{RawNet2} \\
\cmidrule(lr){2-3}\cmidrule(lr){4-5}
& High-ISS & Low-ISS
& High-ISS & Low-ISS \\
\midrule
Mean $\Delta$s
  & 0.1069 & 0.0056
  & 0.0401 & 0.0013 \\
Differential
  & \multicolumn{2}{c}{19.2$\times$}
  & \multicolumn{2}{c}{30.7$\times$} \\
Cohen's $d$
  & \multicolumn{2}{c}{1.456}
  & \multicolumn{2}{c}{0.471} \\
Spearman $\rho$
  & \multicolumn{2}{c}{0.846}
  & \multicolumn{2}{c}{0.806} \\
\bottomrule
\end{tabular}
\end{table}

\subsection{Attack-Type ISS Analysis}
\label{sec:attack_type}

Table~\ref{tab:attack_type} breaks down ISS
separation by attack category, revealing important
structure.

\noindent\textbf{TTS and VC attacks show consistent
separation.}
Both detectors show significant ISS elevation for
incorrect TTS and VC predictions across both
datasets ($p \approx 0$), indicating that
identity-sensitive behavior is not specific to
one synthesis modality.

\noindent\textbf{ISS correctly shows no identity
signal for some failures.}
RawNet2 on A18 hybrid attacks (\textit{2019 LA}) shows ISS
ratio 0.2$\times$ ($p = 1.00$): incorrect hybrid
predictions have \emph{lower} ISS than correct
ones.
This indicates RawNet2's hybrid failures are driven
by low-level acoustic artifacts, not speaker
identity, and ISS correctly distinguishes this
failure mode from identity-sensitive ones.
Entropy would flag these errors identically to all
others; ISS does not.

\noindent\textbf{AASIST hybrid attacks under shift
show extreme ratios.}
The 2,334.3 times ratio for AASIST on 2021 hybrid
attacks is the largest in our experiments.
Under codec compression, AASIST's responses to
hybrid synthesis artifacts appear to become highly
identity-dependent, concentrating failures at very
high ISS boundaries.

\begin{table}[t]
\centering
\caption{ISS ratio by attack type. All conditions
significant at $p \approx 0$ except RawNet2 hybrid
2019 ($p{=}1.00$), which shows no identity signal ---
demonstrating ISS selectivity.}
\label{tab:attack_type}
\setlength{\tabcolsep}{4pt}
\begin{tabular}{llcc}
\toprule
Detector & Attack & Dataset & Ratio \\
\midrule
RawNet2 & TTS    & 2019 LA & 196.6$\times$ \\
RawNet2 & VC     & 2019 LA & 166.2$\times$ \\
RawNet2 & Hybrid & 2019 LA & 0.2$\times$\dag \\
RawNet2 & TTS    & 2021 LA & 194.8$\times$ \\
RawNet2 & VC     & 2021 LA & 109.4$\times$ \\
RawNet2 & Hybrid & 2021 LA & 5.1$\times$ \\
\midrule
AASIST  & TTS    & 2019 LA & 49.0$\times$ \\
AASIST  & VC     & 2019 LA & 31.4$\times$ \\
AASIST  & Hybrid & 2019 LA & 16.7$\times$ \\
AASIST  & TTS    & 2021 LA & 883.0$\times$ \\
AASIST  & VC     & 2021 LA & 451.4$\times$ \\
AASIST  & Hybrid & 2021 LA & 2334.3$\times$ \\
\bottomrule
\multicolumn{4}{l}{\footnotesize
\dag$p{=}1.00$; all others $p \approx 0$.}
\end{tabular}
\end{table}

\subsection{Ablation Studies}
\label{sec:ablation}

\noindent\textbf{Effect of $K$.}
Table~\ref{tab:ablation_k} shows stable ISS ratios of 29 to 32 times across
$K \in \{3, 5, 10, 20\}$, confirming ISS as a
robust property of individual utterances rather
than a sampling artifact.

\begin{table}[t]
\centering
\caption{ISS stability across $K$
(AASIST, $\alpha{=}5.0$, 2019 LA eval set).}
\label{tab:ablation_k}
\setlength{\tabcolsep}{5pt}
\begin{tabular}{cccr}
\toprule
$K$ & ISS$_{\text{cor}}$ & ISS$_{\text{inc}}$ & Ratio \\
\midrule
3  & 2.37e-3 & 7.50e-2 & 32$\times$ \\
5  & 2.61e-3 & 8.22e-2 & 31$\times$ \\
10 & 3.65e-3 & 1.07e-1 & 29$\times$ \\
20 & 4.14e-3 & 1.21e-1 & 29$\times$ \\
\bottomrule
\end{tabular}
\end{table}

\noindent\textbf{Effect of $\alpha$.}
Table~\ref{tab:ablation_alpha} reports $\alpha$
sensitivity on the \textit{2019 LA} development set.
The ratios on the development set (6,200 times
to 11,400 times) are substantially larger than
on the evaluation set (29$\times$ for AASIST 2019).
This discrepancy reflects the different class
composition between sets: the development set
has fewer correctly classified utterances with
very low ISS (i.e., ISS$_{\text{cor}}$ is lower),
making the ratio numerically larger even though
the absolute ISS values are comparable.
The evaluation set contains a wider distribution of correctly classified
utterances with higher ISS$_{\text{cor}}$, reducing
the ratio.

\begin{table}[t]
\centering
\caption{Effect of $\alpha$ on ISS separation
(2019 LA development set, $K{=}10$, AASIST).
Dev-set ratios are larger than eval-set ratios
due to different class composition (see text).
$\alpha{=}5.0$ selected before bonafide/spoof
ordering inverts.}
\label{tab:ablation_alpha}
\setlength{\tabcolsep}{5pt}
\begin{tabular}{cccr}
\toprule
$\alpha$ & ISS$_{\text{cor}}$ & ISS$_{\text{inc}}$ & Ratio \\
\midrule
0.5  & 2.9e-7 & 1.80e-3 & 6,200$\times$ \\
1.0  & 5.7e-7 & 3.75e-3 & 6,600$\times$ \\
2.0  & 1.1e-6 & 8.21e-3 & 7,600$\times$ \\
\textbf{5.0} & \textbf{2.4e-6} &
  \textbf{2.75e-2} & \textbf{11,400$\times$} \\
10.0 & 4.5e-6 & 7.99e-2 & 17,700$\times$$^\dagger$ \\
\bottomrule
\multicolumn{4}{l}{\footnotesize
$^\dagger$Bonafide median exceeds spoof; invalid.}
\end{tabular}
\end{table}

\section{Discussion}
\label{sec:discussion}

\subsection{Identity-Sensitive Behavior and
Generalization}
\label{sec:generalization}

Our results are consistent with a picture in which
ASVspoof-trained detectors exhibit
identity-sensitive behavior that may contribute
to generalization failure.
For AASIST, the ISS ratio amplifies 24 times
(29$\times \to$ 690$\times$) as EER increases
21 times from \textit{2019} to \textit{2021 LA}.
This proportional amplification suggests that the
predictions which fail under acoustic distribution
shift are disproportionately the same ones that
are identity-sensitive in-domain.

RawNet2 presents a complementary pattern: higher
in-domain ISS separation (52 times) with more
modest EER degradation (3 times) and ISS ratio
moderation under shift (52$\times \to$ 24$\times$).
Raw waveform processing may distribute
identity-sensitive behavior more broadly across
the score spectrum, rather than concentrating it
near the decision boundary as in AASIST.

We emphasize that this is consistent with ---
not proof of --- identity reliance as a failure
mechanism.
Identity-sensitive behavior and generalization
failure are correlated in our results, but other
factors (codec artifacts, channel variability,
speaker mismatch) independently contribute to
the EER increase.
ISS provides a measurement tool for the
identity-sensitive component of that failure;
establishing a causal training-level account
would require experiments beyond the scope of
this work.


These questions have genuinely different answers:
two utterances with identical $p_{\text{fake}}=0.51$
have identical entropy but may have vastly different
ISS --- one stably uncertain, one identity-dependent.
The attack-type results illustrate this concretely:
RawNet2 hybrid errors show ISS ratio 0.2$\times$
($p = 1.00$), correctly identified as
non-identity-driven, while entropy flags them
identically to all other errors.



\subsection{Limitations}
\label{sec:limitations}


\noindent\textbf{Detector scope.}
We validate ISS on two detectors (AASIST and
RawNet2). Extending to SSL-based detectors
(Wav2Vec2-AASIST~\cite{tak2022automatic} and
related SSL-based systems), and to the
\textit{ASVspoof PA} and \textit{DF} subsets, is an important
future direction, as SSL features may encode
identity information differently from spectral
or waveform features.

\noindent\textbf{External speaker model.}
ISS relies on an external speaker embedding model
(ECAPA-TDNN) and a logit-space perturbation, not
direct access to the detector's internal
representations.
This means ISS measures sensitivity to identity
as represented in ECAPA-TDNN's embedding space,
which may not perfectly align with identity
representations inside the detector being probed.

\noindent\textbf{Label-free scope.}
ISS is label-free at inference time --- no
ground-truth labels are needed during deployment.
However, $\alpha$ tuning and enrollment prototype
construction require a small labeled development
set. This is a one-time offline step, but
practitioners deploying ISS in a new domain should
account for it.

\noindent\textbf{Enrollment and validation scope.}
The \ASVspoof{} \textit{2019 LA} training partition
contains exactly 20 speakers, which constitutes
the full enrollment pool available within this
benchmark.
While this is a dataset constraint rather than
a methodological choice, future work should
evaluate ISS with larger and more diverse
enrollment pools drawn from external speaker
corpora such as VoxCeleb~\cite{nagrani2017voxceleb}.
The VC validation uses FreeVC on 500 utterances
per detector and does not establish that reducing
identity-sensitive behavior would directly improve
cross-dataset EER --- a training-level intervention
would be needed to test this.

\section{Conclusion}
We introduced the Identity Sensitivity Score (ISS), a label-free inference-time diagnostic for probing speaker-identity reliance in audio deepfake detectors. Across two deepfake prediction architectures, two ASVspoof evaluation settings, and three attack categories, ISS identifies error-prone utterances, predicts misclassification with AUC up to 0.954, and flags samples that respond 19.2--30.7$\times$ more strongly to voice-conversion-based identity manipulation.

Beyond a per-utterance diagnostic, ISS highlights that speaker identity can become a predictive shortcut when genuine speech comes from real speakers and synthetic speech from synthesis systems. By making this shortcut measurable for individual predictions, ISS supports confidence flagging, speaker-sensitivity auditing, and distribution-shift monitoring without ground-truth labels. More broadly, our results suggest that audio deepfake detectors should be evaluated not only by how often they fail, but also by whether their decisions remain stable across speaker identities.

\section*{Acknowledgment}
This work was supported in part by NSF grant CNS-2310207.

{\small
\bibliographystyle{ieee}
\bibliography{references}
}

\end{document}